# Threefold Efficiency Enhancement and Narrowed Nanoparticle Size Distribution in Laser Ablation of Gold in Water by GHz-Burst Irradiation

Maximilian Spellauge[1], Ramon Auer[1], Vincent Taebling[1], Anna R. Ziefuss[2], Daniel J. Förster[1] and Heinz P. Huber[1,3,*]

[1]Laser Center HM, Munich University of Applied Sciences HM, 80335 Munich, Germany

[2]Technical Chemistry I and Center for Nanointegration Duisburg-Essen (CENIDE), University of Duisburg-Essen, 45141 Essen, Germany

[3]New Technologies Research Center, University of West Bohemia, Plzen CZ-30100, Czech Republic

*Corresponding author: heinz.huber@hm.edu

## Abstract

Laser ablation in liquids enables the synthesis of surfactant-free nanoparticles but remains limited in productivity due to intrinsic constraints imposed by the liquid environment. These constraints include nonlinear optical losses, material redeposition, and cavitation bubble-induced shielding. Temporal intensity shaping of the incident laser pulse offers a potential route to mitigate these limitations. Here, ultrashort GHz-burst ablation is applied to laser ablation of gold in water. By distributing the pulse energy into a sequence of picosecond sub-pulses arriving within the nanosecond time window preceding cavitation bubble formation, GHz-burst irradiation enables energy delivery before the onset of bubble-induced shielding. This increases the threshold fluence for nonlinear losses and yields an ablation efficiency enhancement of up to a factor of three compared to single-pulse ablation. Importantly, this efficiency gain is not accompanied by an increase in cavitation bubble size or lifetime. In addition to enhanced efficiency, burst irradiation yields a twofold narrower nanoparticle size distribution. These results demonstrate that GHz-burst ablation is a promising approach to increase productivity while simultaneously improving nanoparticle quality.

## 1. Introduction

Pulsed laser ablation in liquids (LAL) enables the synthesis of surfactant-free nanoparticles from a wide range of materials and avoids the use of chemical precursors [1]. These characteristics make LAL attractive for applications in biomedicine [2] and catalysis [3], as well as for use as functional nanomaterials in additive manufacturing [4].

For LAL to compete with established chemical and physical nanoparticle synthesis routes, productivities on the grams-per-hour scale are required [5,6]. Such productivities have already been demonstrated, with values of 8 g/h for platinum and 4 g/h for gold [7]. However, recent analyses indicate that these values remain up to an order of magnitude below what can be achieved [8], highlighting significant potential for further increases in efficiency and scalability. This gap arises from intrinsic physical constraints imposed by the liquid environment [9].

During propagation through the liquid, Kerr-induced self-focusing can locally increase the intensity, leading to filamentation and, upon reaching the breakdown threshold, to plasma formation via optical breakdown [10]. The generated plasma absorbs and scatters the incident radiation, reducing the fraction of laser energy that reaches the target surface [11].

After energy deposition and the onset of material ablation, the liquid inertia can contribute to the redeposition of ablated material on sub-nanosecond timescales, reducing the net ablation efficiency [8,12,13].

Approximately one to two nanoseconds after pulse impact, cavitation bubble formation sets in [8,14,15], and the expanding vapor cavity can induce shielding of subsequent pulses within the focal volume [7]. The bubble persists for several tens to hundreds of microseconds [15,16]; upon collapse, additional redeposition of material previously trapped within the bubble may occur [17].

Persistent microbubbles and generated nanoparticles can remain in the liquid after bubble collapse [11]. These species absorb and scatter subsequent irradiation, and even a single nanoparticle present in the focal volume can lower the threshold fluence for nonlinear losses by orders of magnitude [18]. In practical systems, liquid flow can remove a substantial fraction of these residual species, although a stationary liquid layer containing nanoparticles may remain near the target surface [11].

These dynamics impose a fundamental temporal constraint in LAL to avoid cavitation-bubble shielding. Pulse energy must be delivered either before the cavitation bubble significantly shields subsequent pulses or after the bubble has collapsed.

Experimental observations support this constraint. Under MHz repetition rate irradiation, ablation efficiency decreases when subsequent pulses arrive during the cavitation bubble lifetime [19], whereas efficiency is recovered when pulses are temporally separated beyond the bubble lifetime or spatially displaced from it [7].

In addition to repetition-rate effects, pulse-duration studies provide further evidence that efficient energy deposition occurs on a timescale preceding significant cavitation bubble development [14]. For ultrashort pulses, nonlinear propagation losses become increasingly relevant, whereas longer nanosecond pulses gradually overlap with the developing cavitation bubble [14]. Double-pulse irradiation with sub-nanosecond delays has further demonstrated that modifying the temporal intensity distribution of the irradiating pulse in this regime can affect nanoparticle formation, for example, by reducing bimodal size distributions, even though ablation efficiency may decrease [20].

We extend this approach using ultrashort GHz-burst irradiation, which delivers the total fluence in a sequence of picosecond pulses separated by hundreds of picoseconds. Based on our prior findings, we hypothesize that repeated interaction with the upward-propagating ablated material may increase ablation efficiency [8]. In liquid processing contexts such as drilling of metallic foils, GHz-burst irradiation has been shown to suppress nonlinear losses in the liquid by reducing the peak intensity and shortening drilling time [21]. However, it remains unclear whether GHz-burst irradiation can enhance nanoparticle productivity in LAL without increasing cavitation bubble shielding or compromising nanoparticle quality.

In this work, we investigate the influence of the temporal intensity distribution by bursts of ultrashort pulses with a 205 ps inter-pulse delay (GHz-burst) for nanoparticle generation by laser ablation in liquids using gold as a model system. The general applicability of GHz-burst processing is evaluated along the complete process chain: (i) energy delivery to the target and threshold fluence for nonlinear losses, (ii) conversion of incident pulse energy into removed material, (iii) cavitation bubble dynamics, (iv) robustness under repeated irradiation, and (v) the resulting nanoparticle size distribution.

## 2. Materials and Methods

### *2.1 Sample preparation*

Polycrystalline Au samples with a purity of 99.99% and a thickness of 1 mm were embedded in an epoxy resin matrix. The samples were sanded and subsequently polished using 9 µm, 3 µm, and 1 µm polycrystalline diamond suspensions. In a final chemical–mechanical polishing step, a suspension of 50 nm $Al_2O_3$ nanoparticles with a few mL $KI/I_2$ solution was used. The resulting average surface roughness of $R_a \approx 20$ nm was determined using confocal microscopy (Ergoplan, Leitz).

### *2.2 Laser system and GHz-burst characterization*

An ultrafast laser source (Pharos, Light Conversion) emitting linearly polarized pulses at 1030 nm with a pulse duration of 10 ps and a repetition rate of 500 Hz was used throughout the study. The beam was near-diffraction-limited ($TEM_{00}$) with $M^2 < 1.1$. The laser source was operated in single-pulse mode or in GHz-burst mode, where the number of sub-pulses per burst $P$ was adjustable between 2 and 25. In the following, a single pulse or a GHz-burst is referred to as an event, where in GHz-burst mode an event comprises $P$ sub-pulses. The single-pulse energy or, for a burst, the total energy (sum of the individual sub-pulse energies) is hereafter denoted as the event energy $E_0$.

The temporal intensity distribution of the GHz-burst was characterized using a fast photodiode (25 ps rise time, ET-3500, EOT) in combination with a 16 GHz bandwidth oscilloscope operating at 40 GS/s (RTP164, Rohde & Schwarz). A representative oscilloscope trace for $P = 5$ is shown in Figure 1a, where the measured inter-pulse delay of 205 ps (4.88 GHz) is indicated. The sub-pulse energies were given by the laser source and increased from pulse 1 to pulse $P$-1, with the final pulse carrying the largest energy fraction.

The energy fraction $E_i$ of each sub-pulse was determined from the photodiode voltage trace by integrating a fixed temporal window around each pulse peak (gray area in Figure 1a). The event energy $E_0$ was obtained by integrating the trace over the full burst duration, covering all sub-pulses. Figure 1b shows the resulting energy fractions $E_i/E_0$ for $P \leq 5$; for $P > 5$ a similar trend was observed.

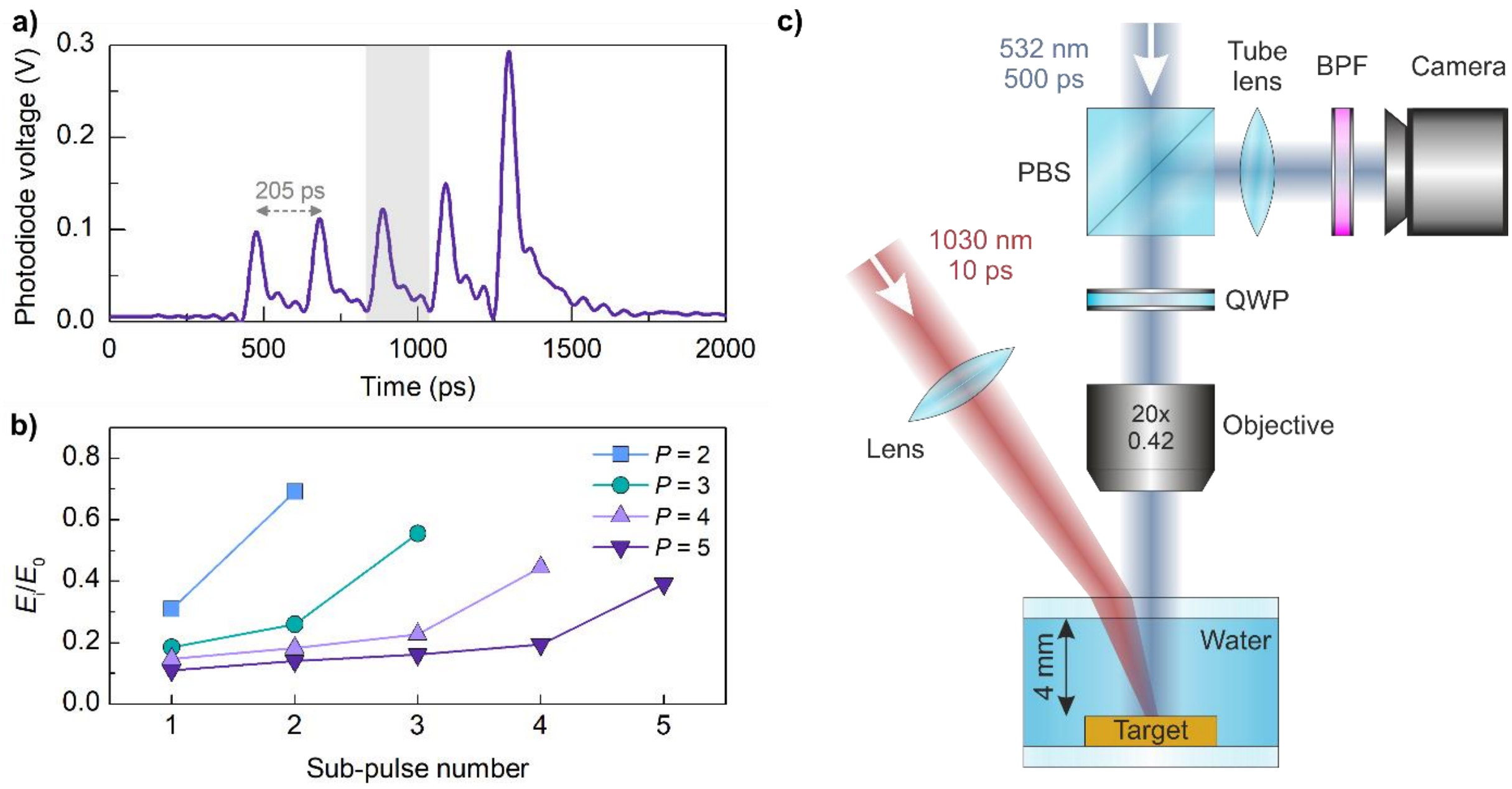

**Figure 1: a** Representative oscilloscope trace of a GHz-burst with $P = 5$ sub-pulses measured using a fast photodiode. The inter-pulse delay of 205 ps is indicated. The shaded region exemplifies the temporal integration window used to determine individual sub-pulse energies. **b** Relative energy fractions $E_i/E_0$ of the individual sub-pulses for $P \leq 5$, obtained by integration of the photodiode voltage trace. **c** Schematic of the experimental configuration. PBS: polarizing beam splitter; QWP: quarter-wave plate; BPF: band-pass filter.

### *2.3 Beam delivery and ablation conditions*

The event energy $E_0$ was adjusted using a half-wave plate and polarizing beam splitter combination. A mechanical shutter (Uniblitz LS6, Vincent Associates) was used to select individual laser events. The beam diameter was adjusted using a beam expander, and the beam was focused on the sample using a plano-convex lens with a focal length of 100 mm (Figure 1c).

Ablation experiments were performed both in air and in water. For ablation in air, the laser beam was directly focused onto the sample surface at an incidence angle of 37°. For ablation in water, the sample was placed in a quartz cuvette, immersed in 4 mm of distilled water, and enclosed with a 1 mm-thick fused silica glass slide (Corning 7980). The beam was incident on the glass cover at an angle of 37°. After refraction at the air-glass and glass-water interfaces, the incidence angle at the sample surface was $\theta = 27°$ (Figure 1c).

The beam waist at the sample was determined using the $D^2$-method [22]. The beam waist radius $w_0$ (minor axis) of the resulting elliptical focal spot was $(13.7 \pm 0.1)$ µm in air and $(16.6 \pm 0.4)$ µm in water ($1/e^2$ intensity definition). The event fluence was calculated as $\Phi_0 = 2 \cdot E_0 \cdot \cos(\theta)/(\pi \cdot w_0^2)$. This fluence corresponds to the peak fluence; the average fluence is lower by a factor of 2 for a Gaussian beam. $E_0$ was measured at the sample position without the cuvette using a power meter (PS10Q, Coherent Inc.); energy fluctuations were below 1%. The transmittance $T$ through the cuvette and water layer was determined according to reference [8], and the incident event fluence was calculated as $\Phi_I = T \cdot \Phi_0$. The transmittance through the fused silica cover slide was measured separately and was constant over the investigated fluence range. For ablation in air, $T = 1$, and hence no transmission correction was applied.

### *2.4 Ablation efficiency determination*

The ablation efficiency was defined as $\eta = V_{abl}/E_I$, where $V_{abl}$ denotes the ablation volume and $E_I$ the incident event energy at the target. The incident event energy was calculated as $E_I = T \cdot E_0$.

The ablation volume was determined using confocal microscopy (Ergoplan, Leitz) with a lateral resolution of 500 nm and an axial resolution of 1 nm. For each parameter set, the ablation volume was averaged over ten individual craters.

The ablated crater volume reflects the net mass removed from the target, as redeposited material within the ablation site is inherently accounted for in the topographical measurement. Under ultrashort-pulse LAL conditions, experimental and theoretical studies have shown that the ablated mass is quantitatively converted into nanoparticles [17,23,24]. The crater volume therefore provides a consistent estimate of nanoparticle mass generation in the present experiments. A direct gravimetric determination was not feasible, as the maximum ablated volume ($\approx 3000$ µm$^3$) corresponds to a removed mass of only $\approx 60$ ng.

### *2.5 Cavitation bubble imaging*

Cavitation bubble imaging was performed using a pump-probe imaging scheme with an electronically delayed second laser pulse at a wavelength of 532 nm and a pulse duration of 500 ps. The delay between the ablation pulse and the illumination pulse was controlled electronically using a delay generator (DG645, Stanford Research Systems). The temporal jitter of the laser source was approximately 1 ns and defines the temporal resolution.

The illumination beam was guided through a 20x microscope objective (Plan Apo 20, Mitutoyo) with a numerical aperture of 0.42 and imaged onto a CMOS camera (pco.pixelfly USB, PCO AG), as shown in Figure 1c.

Images were recorded at defined delay times after the ablation event to capture the temporal evolution of the cavitation bubble. For each delay time, ablation was performed at a pristine sample position to avoid cumulative effects, and two images were recorded. One image $R_0$ was acquired 1 s before pump pulse impact, while the second image $R(\Delta t)$ was captured at the desired delay time $\Delta t$. The relative reflectance change was then calculated pixel-wise according to $\Delta R/R_0 = (R(\Delta t) - R_0)/R_0$. A more detailed description of the imaging setup is provided elsewhere [8].

*2.6 Nanoparticle size distribution analysis*

A droplet of distilled water containing 100 µM NaCl was placed onto the Au surface and covered with a cover slip. The NaCl stabilizes the generated nanoparticles electrostatically and prevents agglomeration. The resulting liquid layer had a height of approximately 1 mm.

Within the droplet, ablation was performed at nine positions arranged in a square pattern with a spacing of 500 µm between adjacent positions. Each position was irradiated by ten events.

After ablation, the cover slip was removed and the droplet containing the nanoparticles was collected using a pipette. The suspension was drop-cast onto a transmission electron microscopy grid (EMR Lacey Carbon support film on a nickel 400 square mesh, diameter 2 mm) and dried under ambient conditions. The total time between ablation and complete drying was approximately 1 h.

Transmission electron microscopy (JEM-1400Plus, JEOL) operated at an acceleration voltage of 120 kV was used to determine the nanoparticle size.

## 3. Results and Discussion

*3.1 Transmittance*

To investigate nonlinear losses under GHz-burst excitation, Figure 2 shows the transmittance $T$ through the cover glass and 4 mm water layer as a function of the event fluence $\Phi_0$ for different sub-pulse numbers $P$ within the GHz-burst. For all sub-pulse numbers, the transmittance remains approximately constant with increasing fluence and then decreases abruptly, marking the threshold fluence for the onset of nonlinear losses $\Phi_{NL}$ [25]. $\Phi_{NL}$ increases with sub-pulse number, rising from 9 J/cm$^2$ for $P = 1$ to 18 J/cm$^2$ for $P = 3$ (see inset in Figure 2). Starting from $P = 4$, no nonlinear losses are observed within the accessible fluence range up to 22 J/cm$^2$.

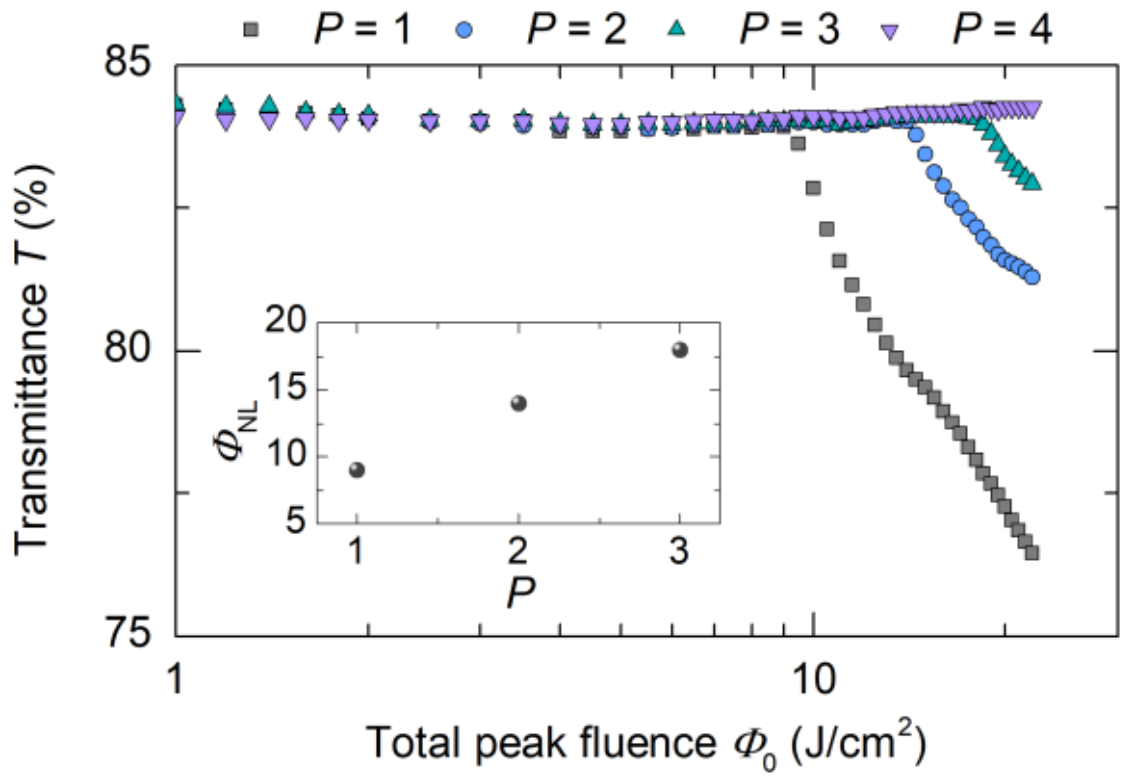

**Figure 2:** Transmittance $T$ through the water layer as a function of event fluence $\Phi_0$ for different sub-pulse numbers $P$ within the GHz-burst. The inset shows the threshold fluence for nonlinear losses $\Phi_{NL}$ as a function of sub-pulse number $P$.

In the present GHz-burst configuration, the pulse energies are not uniformly distributed, with the last pulse carrying the highest energy fraction (Figure 1a). Multiplying $\Phi_{NL}$ by the energy fraction of the last pulse (Figure 1b) yields values of approximately 9.8 J/cm$^2$ for $P = 2$ and 9.9 J/cm$^2$ for $P = 3$, close to the single-pulse threshold of 9 J/cm$^2$. This is consistent with nonlinear losses being initiated by the most energetic sub-pulse. Free seed electrons generated by preceding pulses could, in principle lower the threshold for subsequent pulses [26]. However, in water these electrons undergo rapid solvation and recombination on sub-picosecond timescales [27], which is substantially shorter than the 205 ps inter-pulse delay. They therefore do not influence subsequent pulses.

Distributing the fluence across multiple sub-pulses thus increases the total fluence that can be delivered to the target without inducing nonlinear losses in the liquid.

*3.2 Ablation efficiency*

Figure 3 shows the single-event ablation efficiency as a function of the number of sub-pulses per burst P for different incident event fluences $\Phi_I$ during ablation in water (left) and air (right) under otherwise identical conditions.

In water, the single-pulse efficiency amounts to approximately 0.2 µm$^3$/µJ at 6 J/cm$^2$ and approximately 0.8 µm$^3$/µJ at both 12 J/cm$^2$ and 18 J/cm$^2$. At 6 J/cm$^2$, the efficiency decreases with increasing $P$, and no ablation is observed beyond $P = 5$. At 12 J/cm$^2$ and 18 J/cm$^2$, the efficiency initially decreases at $P = 2$, consistent with previous observations for double-pulse ablation of gold in water [20], followed by an increase. For 12 J/cm$^2$, a maximum is reached at $P = 4$, whereas for 18 J/cm$^2$ the maximum efficiency exhibits a plateau spanning from $P = 5$ to $P = 10$. At 18 J/cm$^2$, a maximum efficiency of 2.3 µm$^3$/µJ is observed in this plateau, exceeding the corresponding single-pulse value by a factor of approximately 3. For larger numbers of sub-pulses, the efficiency decreases again.

In contrast, in air the efficiency decreases monotonically from the single-pulse value of approximately 2 µm$^3$/µJ, which is about a factor of two higher than the corresponding single-pulse efficiency in water. Starting from $P = 5$ the ablation efficiency approaches zero. The present experiments are conducted in a regime where the fluence of each sub-pulse exceeds or is close to the ablation threshold. In this regime, the observed efficiency decrease is attributed to redeposition of ablated materials and shielding [28–30]. This behavior contrasts with GHz-burst ablation at sub-threshold fluence per sub-pulse, where efficiency enhancement is typically observed for large sub-pulse numbers exceeding 1000 [30,31].

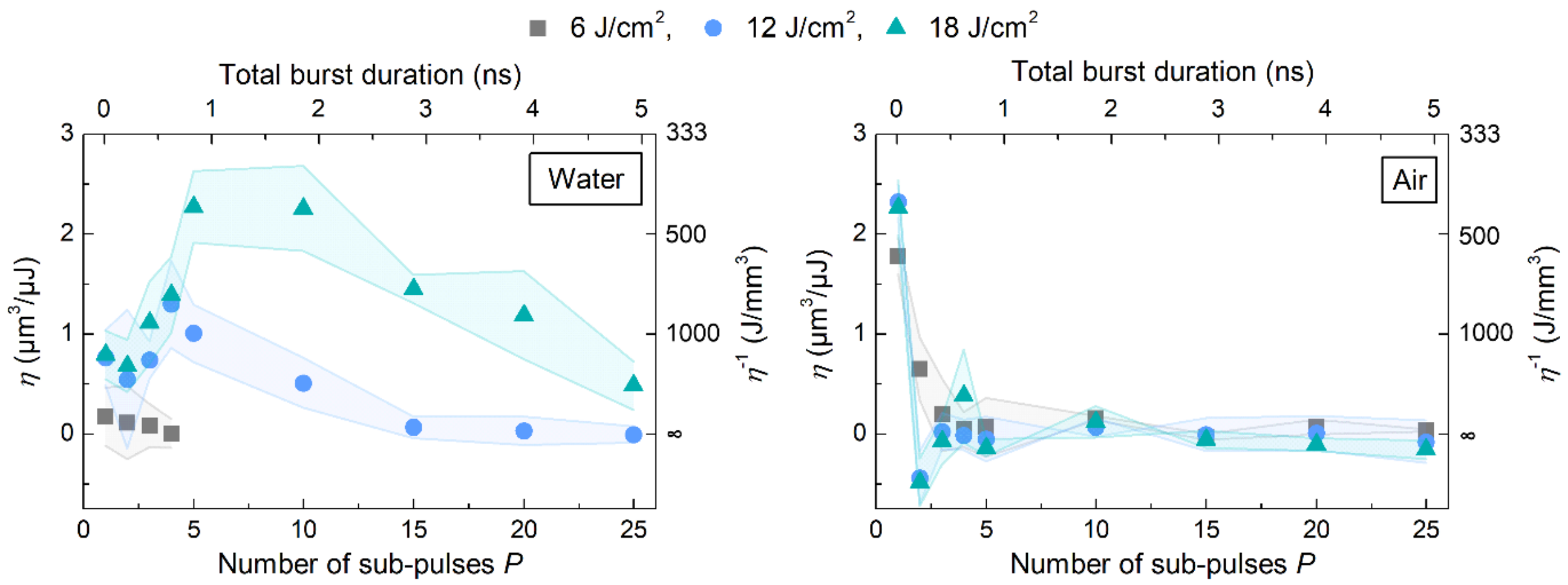

**Figure 3:** Single-event ablation efficiency $\eta$ as a function of the number of sub-pulses per burst $P$ for various incident event fluences during laser ablation in water (left) and air (right). The corresponding total burst duration is given on the top x-axis, and the inverse ablation efficiency $\eta^{-1}$, corresponding to the specific energy for ablation, is shown on the right y-axis.

The efficiency enhancement in water is not observed in air, indicating that the increase is specific to laser ablation in water. Since the efficiency is plotted versus the incident event fluence corrected for transmission losses in the liquid (Section 3.1), the observed enhancement cannot be attributed to reduced transmission losses in the liquid for GHz-burst ablation. The approximately threefold efficiency drop from 2.3 $\mu m^3/\mu J$ in air to 0.8 $\mu m^3/\mu J$ in water in the single-pulse regime has previously been attributed to redeposition of ablated material under liquid confinement [8,12,13]. Notably, at an incident event fluence of 18 $J/cm^2$ the maximum efficiency achieved in water under GHz-burst ablation (2.3 $\mu m^3/\mu J$) is equal to the single-pulse efficiency measured in air (2.3 $\mu m^3/\mu J$). This observation is consistent with a reduction of liquid-confinement-induced losses, in particular redeposition. Overall, single-event GHz-burst ablation of Au in water enables an efficiency enhancement of up to a factor of 3 compared to single-pulse ablation.

To place the achieved ablation efficiency in perspective, the inverse ablation efficiency is considered, corresponding to the specific energy needed for ablation. For the maximum efficiency of 2.3 $\mu m^3/\mu J$ observed under GHz-burst irradiation, this yields a value of approximately 440 $J/mm^3$ (Figure 3, left). Reported record values for the specific energy for ablation of Au in water amount to 3333 $J/mm^3$, achieved using a 3 ps laser source in combination with a polygon scanner operated at a scanning speed of 484 m/s to avoid cavitation bubble shielding [7]. In comparison, nanosecond LAL at a pulse duration of 1 ns yields 1111 $J/mm^3$ [32]. This pulse duration is long enough to avoid nonlinear losses in the liquid and short enough to prevent shielding of the trailing edge of the pulse by the cavitation bubble generated by the leading edge [14]. Thus, GHz-burst ablation reduces the specific energy for ablation by nearly a factor of 8 compared to the ultrafast LAL benchmark and by a factor of approximately 2.5 compared to the 1 ns benchmark.

*3.3 Cavitation bubble dynamics*

While GHz-burst increases ablation efficiency in water, its applicability to scalable processing requires that this gain is not mitigated by increased cavitation bubble shielding. An increased cavitation bubble size or lifetime would reduce the feasible repetition rate at a given scanning speed and thereby reduce the deliverable laser energy to the target [7].

Figure 4a shows images of the relative reflectance change for single-pulse ablation ($P = 1$, top row) and GHz-burst ablation ($P = 5$, bottom row) in water at an incident event fluence of 18 $J/cm^2$, where the largest efficiency enhancement was observed (see Figure 3).

At negative delay times ($\Delta t = -5$ ns), an emission signal is observed for GHz-burst ablation, while no change is visible for single-pulse ablation. This signal arises from plasma emission generated by the pump pulse, which is detected due to the finite temporal gate width of the camera. The increased emission intensity under GHz-burst irradiation is consistent with previous observations for double-pulse ablation, where enhanced plasma emission has been reported for metals such as Al and Ti [33].

For single-pulse irradiation, a pronounced stripe of decreased reflectance is observed along the laser propagation path, which is absent for GHz-burst irradiation. This feature is attributed to optical breakdown within the liquid, leading to plasma formation and absorption of the probe beam, resulting in a reduced reflectance along the beam path [25].

At $\Delta t = 100$ ns, a shockwave is observed as a dark ring surrounding the interaction region, while a central dark region appears corresponding to the cavitation bubble. At $\Delta t = 1$ µs, the cavitation bubble has expanded further. The bubble reaches its maximum size at approximately 20 µs, followed by its collapse. At $\Delta t = 40$ µs, the bubble is still visible for $P = 1$, whereas it has already collapsed for $P = 5$. At later times ($\Delta t = 100$ µs), persistent microbubbles are observed following cavitation bubble collapse.

To quantify these observations, the temporal evolution of the cavitation bubble radius is shown in Figure 4b. The inset highlights the early-time dynamics, where the onset of cavitation bubble formation is defined as the time at which the bubble radius exceeds the final ablation crater radius [14]. Using this definition, onset times of approximately 3 ns for $P = 1$ and 2 ns for $P = 5$ are obtained, consistent with previous observations of cavitation bubble formation on a nanosecond timescale [14].

Following this onset, the bubble expands and reaches a maximum radius of approximately 225 µm for $P = 1$ and 200 µm for P = 5 at about 20 µs. The bubble then collapses at approximately 45 µs for $P = 1$ and at approximately 40 µs for $P = 5$. After the primary collapse, bubble rebounds are visible up to approximately 80 µs for $P = 1$ and up to approximately 70 µs for $P = 5$, beyond which no cavitation bubble is detected in either case.

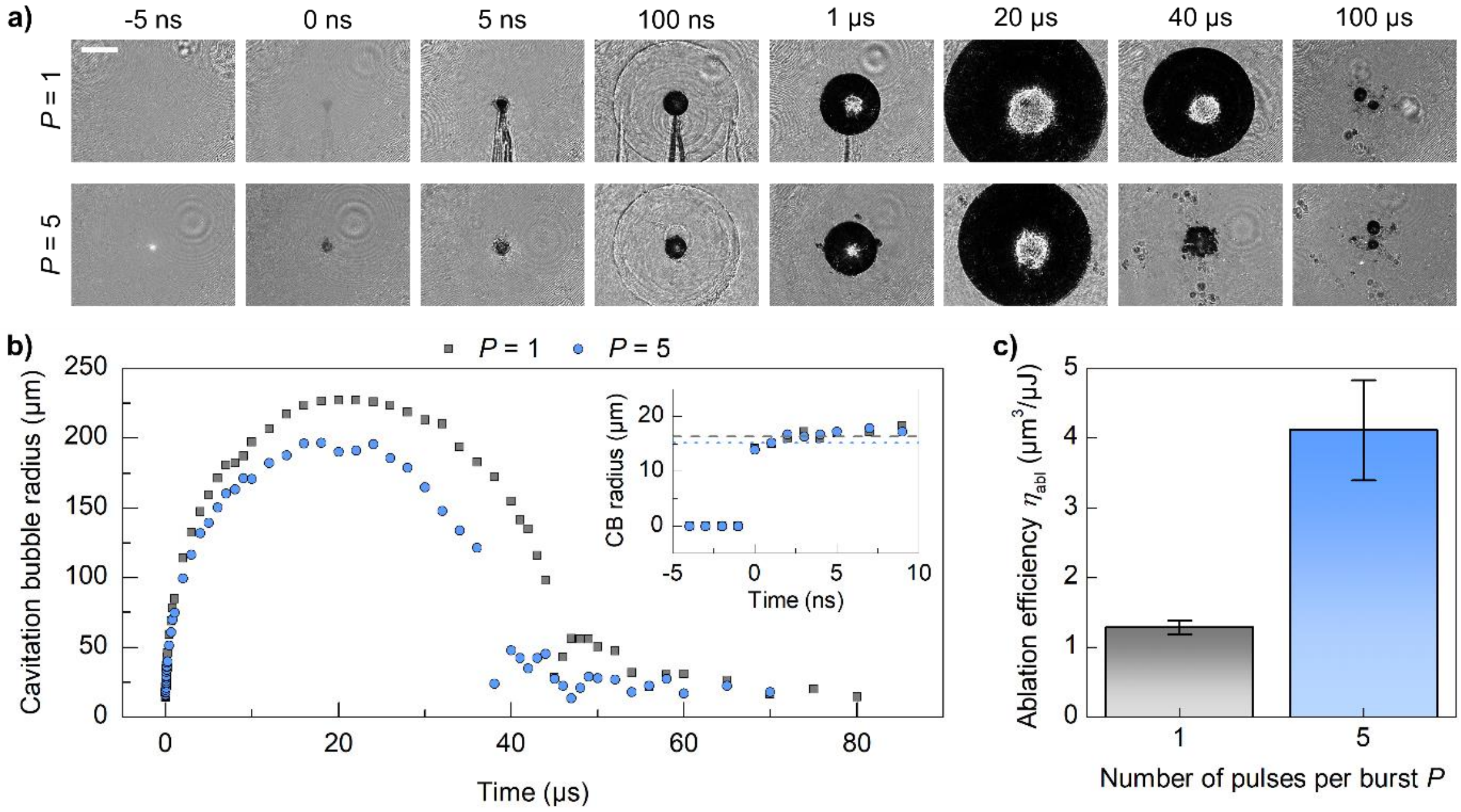

**Figure 4: a** Images showing the relative reflectance change at the indicated delay times for ($P = 1$, top row) and GHz-burst ablation ($P = 5$, bottom row) in water at an incident event fluence of 18 J/cm$^2$. The incidence direction of the laser is from the bottom of the images, the scale bar (top left) corresponds to 100 µm and applies to all images. **b** Temporal evolution of the cavitation bubble radius for $P = 1$ and $P = 5$ under identical conditions. The inset shows the early-time cavitation bubble radius (CB). The dashed black and dotted blue lines indicate the final ablation radius for $P = 1$ and $P = 5$, respectively. **c** Ablation efficiency for $P = 1$ and $P = 5$ at an incident event fluence of 18 J/cm$^2$ under repeated irradiation of the same position with 10 events per position separated by 10 s.

The cavitation bubble energy at maximum expansion can be estimated from $E_{CB} = 4/3 \cdot \pi \cdot r_{max}^3 \cdot (p_0 - p_v)$, where $p_0 = 0.1$ MPa is the ambient pressure and $p_v = 2230$ Pa the vapor pressure of water [25]. Using the measured maximum cavitation bubble radii $r_{max}$ of approximately 225 µm for $P = 1$ and 200 µm for $P = 5$ the bubble energies amount to approximately 4.7 µJ and 3.3 µJ, respectively. This corresponds to a reduction of about 30%, indicating that a smaller fraction of the incident event energy is converted into cavitation bubble energy for GHz-burst ablation.

The higher bubble energy observed for single-pulse irradiation is attributed to optical breakdown within the liquid, as evidenced by the pronounced stripe of reduced reflectance along the beam propagation path in Figure 4a. This process leads to additional energy deposition into the liquid compared to GHz-burst irradiation, where such nonlinear losses are absent, thereby increasing the available energy for cavitation bubble formation [34].

The onset time for cavitation bubble formation of approximately 2 ns observed for GHz-burst irradiation is comparable to the maximum burst duration for which the maximum efficiency enhancement is observed (Figure 3). For larger numbers of sub-pulses ($P > 10$), the total burst duration exceeds this timescale. This suggests that for $P > 10$, sub-pulses arriving after cavitation bubble formation may be partially shielded by the already forming cavitation bubble, resulting in a reduction in ablation efficiency.

Overall, GHz-burst ablation decreases cavitation bubble size and lifetime and therefore does not impose additional constraints on scalability under the present conditions

*3.4 Repeated irradiation*

To mimic scalable LAL conditions in which multiple irradiation events interact with the same target location, the ablation efficiency was evaluated under repeated irradiation of the same position (Figure 4c). A temporal delay of 10 s between events was chosen to exclude shielding by cavitation bubbles, which fully disperse on this timescale (Figure 4b). However, nanoparticles generated during ablation may remain in the interaction region and contribute to shielding [11].

Single-pulse ($P = 1$) and GHz-burst ($P = 5$) ablation in water were compared at an incident event fluence of 18 J/cm$^2$, where the largest efficiency enhancement was observed (Figure 3). For single-pulse ablation, the single-event efficiency of approximately 0.8 µm$^3$/µJ increases to approximately 1.3 µm$^3$/µJ after ten events (Figure 4c), corresponding to an efficiency increase of about 60%. For GHz-burst ablation, the single-event efficiency of approximately 2.3 µm$^3$/µJ increases to approximately 4.1 µm$^3$/µJ, corresponding to an efficiency increase of about 80%.

Within the experimental uncertainty, the relative increase in ablation efficiency from a single event to ten events per position is similar for single-pulse and GHz-burst irradiation. This indicates that GHz-burst ablation does not introduce additional incubation effects under the present experimental conditions. Given the 10 s delay between events, heat accumulation can be excluded [35,36]. The observed efficiency increase must therefore originate from structural modifications of the target surface, such as increased surface roughness or defect density, which can enhance absorption [37,38] and reduce the ablation threshold for subsequent pulses [39]. Notably, the efficiency ratio between GHz-burst and single-pulse ablation increases from approximately 3 in the single-event regime to approximately 3.2 under repeated irradiation, indicating that the GHz-burst efficiency increase is amplified under multi-event ablation.

*3.5 Nanoparticle size distribution*

The nanoparticle size distribution obtained after ablation with ten events per position at an incident event fluence of 18 J/cm$^2$ is shown in Figure 5 for $P = 1$ and $P = 5$. The normalized, number-weighted probability density function (PDF) and cumulative distribution function (CDF) are displayed up to nanoparticle diameters of 30 nm, which encompasses the dominant fraction of the generated nanoparticles (> 98%). Larger nanoparticles with diameters up to approximately 150 nm were observed for both single-pulse and GHz-burst ablation but are not shown for clarity. The overall size distribution, comprising a dominant fraction below 10 nm and a broader minor fraction extending into the several-10 nm range (and up to approximately 150 nm), is consistent with two established nanoparticle generation pathways: evaporation from the target surface followed by condensation produces small primary particles, whereas hydrodynamic instability-driven breakup of a released molten layer results in larger secondary nanoparticles [40,41].

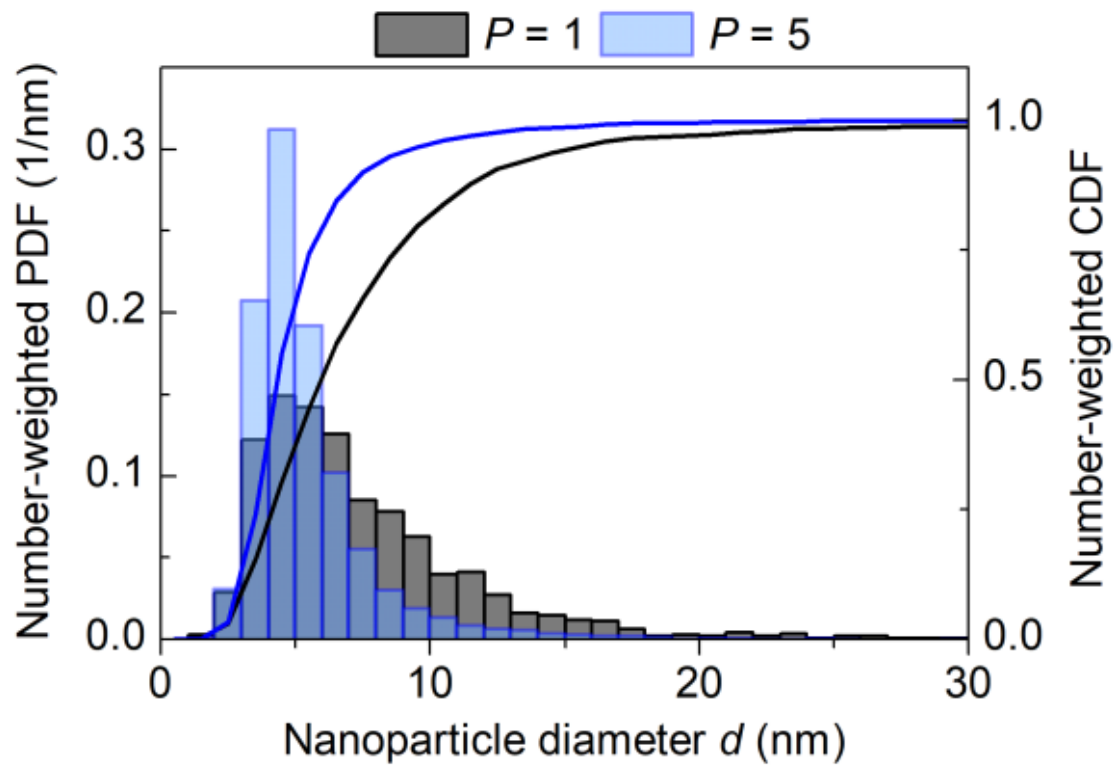

**Figure 5:** Number-weighted probability density function (PDF, bars) and cumulative distribution function (CDF, lines) of nanoparticles generated after ablation with an incident event fluence of 18 $J/cm^2$ in water with 10 events per position for single-pulse ablation ($P = 1$) and GHz-burst ablation with five sub-pulses ($P = 5$). The particle size distributions were obtained by transmission electron microscopy based on 2287 nanoparticles for $P = 1$ and 6920 nanoparticles for $P = 5$.

Both ablation modes predominantly produce nanoparticles with diameters below 15 nm. However, GHz-burst ablation yields a larger fraction of particles below 10 nm (96% for $P = 5$ compared to 83% for $P = 1$). In addition, the particle size distribution obtained under GHz-burst ablation is narrower, as reflected by a reduced interquartile range of 2 nm for $P = 5$ compared to 5 nm for $P = 1$.

A similar effect has been observed in double-pulse ablation in liquids, where sub-nanosecond delayed pulses reduce the fraction of large secondary nanoparticles by approximately 9 wt% [20].

In the GHz-burst regime, subsequent sub-pulses arrive within the sub-nanosecond time window during which secondary nanoparticles are formed [40,41]. Two effects may therefore contribute. First, the repeated interaction may interfere with the generation mechanisms of the secondary nanoparticles, suppressing their formation. Second, already formed secondary nanoparticles may be fragmented [42] by subsequent sub-pulses.

## 4. Conclusion

Laser ablation in liquids is widely used for the synthesis of surfactant-free nanoparticles, but its efficiency is often limited by the presence of the liquid environment, which introduces nonlinear losses, material redeposition, and cavitation bubble shielding. Improving energy conversion into removed material without increasing cavitation-bubble shielding while maintaining nanoparticle size distribution is therefore central for scalable nanoparticle generation.

In this work, temporal intensity distribution shaping using GHz-bursts was investigated as a strategy to increase ablation efficiency and nanoparticle quality for laser ablation of gold in water. GHz-burst ablation increases the threshold fluence for nonlinear losses in the liquid by distributing the peak intensity across sub-pulses, thereby allowing higher pulse energies to be delivered to the target without inducing nonlinear losses. By applying the GHz-burst sub-pulses before the onset of the cavitation bubble, an ablation efficiency enhancement of up to a factor of three is achieved compared to single-pulse ablation. We conclude that this reduces redeposition of ablated material.

When the total burst duration exceeds the cavitation bubble onset time of approximately 2 ns, the ablation efficiency decreases, indicating that sub-pulses arriving after bubble formation are partially shielded. This identifies the temporal window prior to cavitation bubble formation as the regime accessible for temporal intensity shaping.

Importantly, this efficiency increase is not accompanied by an increase in cavitation bubble size or lifetime. The maximum bubble radius decreases from 225 µm to 200 µm, corresponding to an approximately 30% lower bubble energy, and the primary collapse occurs about 5 µs earlier. Thus, GHz-burst irradiation does not impose additional constraints related to cavitation-bubble shielding. Furthermore, GHz-burst irradiation yields smaller nanoparticles with an approximately twofold narrower size distribution.

These findings demonstrate that GHz-burst ablation in water improves ablation efficiency and nanoparticle size distribution by enabling optimized energy delivery prior to cavitation-bubble formation, thereby suppressing redeposition.

**CRediT author contribution statement**

**Maximilian Spellauge:** Conceptualization, Methodology, Investigation, Formal analysis, Data curation, Writing – original draft, Writing – review & editing, Supervision. **Ramon Auer:** Investigation, Formal analysis, Writing – review & editing. **Vincent Taebling:** Investigation, Formal analysis, Writing – review & editing. **Anna R. Ziefuss:** Investigation, Writing – review & editing. **Daniel J. Förster:** Supervision, Writing – review & editing. **Stephan Barcikowski:** Supervision, Writing – review & editing. **Heinz P. Huber:** Conceptualization, Supervision, Funding acquisition, Writing – review & editing.

**Acknowledgements**

The authors gratefully acknowledge financial support by the Deutsche Forschungsgemeinschaft (DFG) under projects 428315411 and 562785215. Furthermore, we would like to thank Jurij Jakobi from University of Duisburg-Essen, for the TEM measurements.

**Declaration of competing interest**

The authors declare no competing interests.

**Declaration of generative AI and AI-assisted technologies in the manuscript preparation process**

During the preparation of this work the authors used ChatGPT for language and grammar polishing. After using this tool, the authors reviewed and edited the content as needed and take full responsibility for the content of the published article.

**Data Availability Statement**

The data that support the findings of this study are openly available in Zenodo at 10.5281/zenodo.20155376.